\documentclass[11pt,a4paper]{article}
\usepackage{amssymb, amsmath, amsfonts, amsthm, graphicx}
\usepackage{jcappub}
\usepackage{multirow}

\newcommand{\e}{{\rm e}}

\title{Kantowski-Sachs Einstein-Aether Scalar Field Cosmological Models}

\author[a]{R. J. van den Hoogen,}
\author[b]{A. A. Coley,}
\author[b]{B. Alhulaimi,}
\author[a]{S. Mohandas,}
\author[a]{E. Knighton,}
\author[b]{S. O'Neil}

\affiliation[a]{Department of Mathematics, Statistics and Computer Science, \\
St. Francis Xavier University,\\
2323 Notre Dame Avenue, Antigonish, N.S., Canada}
\affiliation[b]{Department of Mathematics, \\
Dalhousie University,\\
6316 Coburg Road, Halifax, N.S., B3H 4R2, Canada}

\emailAdd{rvandenh@stfx.ca}
\emailAdd{aac@mathstat.dal.ca}
\emailAdd{bs748397@dal.ca}
\emailAdd{x2015jvq@stfx.ca}
\emailAdd{x2016mrk@stfx.ca}
\emailAdd{saoneil@live.com}

\abstract{
A class of positive curvature spatially homogeneous but anisotropic cosmological models within an Einstein-aether gravitational framework are investigated.  The matter source is assumed to be a scalar field which is coupled to the expansion of the aether field through a generalized exponential potential.   The evolution equations are expressed in terms of expansion-normalized variables to produce an autonomous system of ordinary differential equations suitable for a numerical and qualitative analysis.  An analysis of the local stability of the equilibrium points indicates that there exists a range of values of the parameters in which there exists an accelerating expansionary future attractor.  In general relativity, scalar field models with an exponential potential $V=V_0e^{-2k\phi}$ have a late-time inflationary attractor for $k^2<\frac{1}{2}$; however, it is found that the existence of the coupling between the aether and scalar fields allows for arbitrarily large values of the parameter $k$.}

\arxivnumber{1234.5678}

\begin{document}
\maketitle


\section{Introduction}

\subsection{Lorentz-violating Scalar Field Cosmological Models}

Quantum gravity may predict a preferred rest frame at the microscopic level in vacuum necessitating a violation of Lorentz invariance \cite{Liberati:2013xla,Jacobson:2008aj}.  Further, in cosmology there is a natural frame associated with the cosmic microwave background, and therefore it is possible the assumption of Lorentz invariance may need to be relaxed even at the macroscopic level.   Fortunately, a gravitational theory that violates Lorentz invariance while maintaining general covariance has been proposed within the framework of general relativity (GR), which could possibly address these concerns from both the micro and macro perspectives.

Einstein-aether theory \cite{Jacobson:2000xp,Jacobson:2008aj,Jacobson:2004ts,Jacobson:2010mxa,Jacobson:2010mxb,Garfinkle:2007bk,Garfinkle:2011iw} consists of GR coupled, at second derivative order, to a dynamical timelike unit vector field, the aether. This aether vector field spontaneously breaks Lorentz invariance by selecting a preferred direction at each spacetime point while maintaining local rotational symmetry.   Therefore, it is only the boost sector of the Lorentz symmetry that is actually broken. In this effective field theory, the local space-time structure is determined by a dynamical time-like vector field, $u^a$ (the aether), together with a metric tensor, $g_{ab}$.

In standard Einstein-aether theory, the violation of Lorentz invariance is assumed to be within the gravity sector only.  Since the matter sector is assumed to be coupled only to the metric, the matter sector will remain Lorentz invariant.  Therefore, a simple extension of the standard Einstein-aether theory is to permit Lorentz violations within the matter sector as well. Such extensions have been considered in which the matter is described by a scalar field that couples to the aether via the expansion and shear within the scalar field potential \cite{Donnelly:2010cr,Barrow:2012qy,Solomon:2013iza,Solomon:2015hja,Alhulaimi:2017,Alhulaimi:2017ocb}.  Thus, this extension of the standard Einstein-aether theory opens up the interesting possibility of new cosmological dynamics that depend nontrivially on the expansion and the shear. Indeed, if the scalar field is identified with the inflaton, the aether coupling could potentially modify inflationary dynamics \cite{Alhulaimi:2017ocb}.

In Einstein-aether theory with inflation caused by a scalar field there exists the possibility that inflation is affected by the aether through a direct coupling of the scalar field with the aether field through the expansion and/or shear of the aether vector field. This is in contrast to the situation in a purely metric theory like general relativity, where such quantities cannot be constructed from the metric in a covariant way \cite{Donnelly:2010cr}. Explicit couplings have been studied for spatially homogeneous and isotropic models in several papers \cite{Donnelly:2010cr,Barrow:2012qy,Solomon:2013iza,Solomon:2015hja,Sandin:2012gq,Alhulaimi:2017,Alhulaimi:2017ocb}.  In particular, Donnelly and Jacobson \cite{Donnelly:2010cr} found that the scalar field/aether field coupling can either slow down or speed up the evolution of inflation within a chaotic inflationary scenario.  Barrow \cite{Barrow:2012qy} found solutions in which the coupling parameter enables inflation in situations in which it would not otherwise occur when the scalar field/aether field coupling was exponential in nature. Sandin {\em et al.}, \cite{Sandin:2012gq} using a similar coupling ansatz, found a significant change in the future asymptotic behaviour for large values of the coupling parameter. Alhulaimi \cite{Alhulaimi:2017,Alhulaimi:2017ocb} investigated a scalar field/aether field coupling that also included a coupling to the shear of the aether vector field.  A coupling prescription in which the aether parameters $c_i$ in equation \eqref{K} are functions of the scalar field results in the possibility of having inflation without a scalar field potential,  e.g., with a massless scalar field \cite{Kanno:2006ty}.

We shall investigate a class of locally spherically symmetric spatially homogeneous Einstein-aether cosmological models containing a scalar field using a $1+3$ formalism \cite{wainwright_ellis2005,vanElst:1996dr}. It is assumed that the scalar field is coupled to the aether field through the scalar field potential.  In particular, we explore the potential impact of the scalar field/aether field coupling on the potential isotropization and accelerated expansion within these Kantowski-Sachs scalar field models having a generalized exponential potential. An analysis of the Kantowski-Sachs perfect fluid models within an Einstein-aether paradigm was presented \cite{Coley:2015qqa}.

\subsection{Einstein-aether Gravity with a Scalar Field}

The action under consideration contains a Lagrangian for Einstein-aether (\textsc{AE}) gravity together with a Lagrangian for the matter field  (\textsc{M})
\begin{equation}
S=\int d^{4}x\sqrt{-g}\left[\frac{1}{8\pi G} {\mathcal L}^{\textsc{AE}}+{\mathcal L}^{\textsc{M}} \right] .\label{action}
\end{equation}
The Lagrangian for Einstein-aether gravity (see \cite{Jacobson:2010mxa,Jacobson:2000xp,Jacobson:2010mxb,Garfinkle:2011iw,Garfinkle:2007bk,Jacobson:2008aj,Jacobson:2004ts,Donnelly:2010cr,Solomon:2013iza,Coley:2015qqa,Alhulaimi:2017ocb} for details) is
\begin{equation}
{\mathcal L}^{\textsc{AE}}=\frac{1}{2}R - K^{ab}_{\phantom{ab} cd}\nabla_{a}u^c\nabla_{b}u^d  + \lambda(u^au_a+1), \label{Lagrangian_AE}
\end{equation}
where
\begin{equation}
K^{ab}_{\phantom{ab}{cd}}  \equiv  c_1 g^{ab} g_{cd} + c_2\delta_{c}^{a} \delta_{d}^{b}+c_3\delta_{d}^{a}\delta_{c}^{b}+ c_4 u^{a} u^{b} g_{cd},\label{K}
\end{equation}
and where $\lambda$ is the Lagrange multiplier.  The parameters $c_i$ defined here are the same as those used in \cite{Coley:2015qqa,Alhulaimi:2017ocb} which are equal to half of the values of the $c_i$ employed in \cite{Garfinkle:2011iw,Garfinkle:2007bk} with an opposite sign for the $c_4$.  The metric signature is assumed to be $+2$.

The Lagrangian for the matter is assumed to be that of a single scalar field having a potential that is assumed to be a function of the scalar field, the expansion scalar, and the shear scalar of the aether vector field and therefore has the form:
\begin{equation}
L^\textsc{{M}}= -\frac{1}{2} g^{a b } \nabla_{a}\phi \nabla_{b}\phi - V(\phi,\theta,\sigma^2),
\end{equation}
where $\theta=\nabla_a u^a$ is the expansion scalar and $\sigma^2=\frac{1}{2}\sigma_{ab}\sigma^{ab}$ is the shear scalar.  In theory the scalar field potential $V$ could be a function of any scalar that can be constructed from the aether vector and its derivatives, however, only $\theta$ and $\sigma^2$ are non-trivial in the Kantowski-Sachs models.

Varying \eqref{action} with respect to the metric $g^{ab}$, the aether vector $u^a$, the scalar field $\phi$, and the multiplier $\lambda$ respectively yields the Einstein-aether, Klein-Gordon and the normalization field equations
\begin{eqnarray}
G_{ab}   &=& T_{ab}^{\textsc{U}} + 8\pi G \,T_{ab}^{\textsc{M}},\\
-2\lambda u_a &=&  \frac{\delta {\mathcal L}^{\textsc{U}}}{\delta u^a}+8\pi G\,\frac{\delta {\mathcal L}^{\textsc{M}}}{\delta u^a},\label{deltaL-deltau}\\
\nabla^a\nabla_a\phi-V_{\phi}&=&0.\label{GeneralKleinGordoneqaution}\\
u^a u_a &=& -1,
\end{eqnarray}
When the contributions from the Lagrange multiplier $\lambda$ via a contraction of equation \eqref{deltaL-deltau} with $u^a$ are taken into account, the expression for the effective energy momentum tensor due to the aether vector field becomes
\begin{eqnarray}
T_{ab}^{\textsc{U}} &=& 2\nabla_{c}\Bigl(J_{(a}^{\phantom{a}c}u^{\phantom{c}}_{b)} - J^c_{\phantom{c} (a}u^{\phantom{c}}_{b)} -J_{(ab)}u^c \Bigr) \nonumber\\
            && +2c_1\Bigl((\nabla_{a} u^c)(\nabla_{b} u_c) - (\nabla^c u_a)(\nabla_c u_b)\Bigr) -2c_4\dot{u}_a\dot{u}_b \nonumber\\
            && -2\left( u^d\nabla_c J^{c}_{\phantom{c}d}+c_4\dot{u}_c\dot{u}^c   \right)u_au_b - g_{ab} \Bigl(K^{cd}_{\phantom{cd}ef} \nabla_{c} u^e \nabla_{d} u^f\Bigr),\label{T_ab_AE}
\end{eqnarray}
where
\begin{eqnarray}
J^{a}_{\phantom{a}b}  &=& -K^{ac}_{\phantom{ac}bd} \nabla_{c} u^d,\\
\dot{u}^a &=& u^b\nabla_b u^a.
\end{eqnarray}
Similarly, the effective energy momentum tensor due to the scalar field becomes
\begin{eqnarray}
T_{ab}^{\textsc{M}} &=&  \nabla_a \phi \nabla_b \phi - \left(\frac{1}{2} \nabla_c\phi \nabla^c \phi + V \right)g_{ab}
+ \theta V_\theta g_{ab}+\dot{V}_\theta h_{ab} \nonumber\\
&& +\left(\theta V_{\sigma^2}+\dot{V}_{\sigma^2}\right)\sigma_{ab}+V_{\sigma^2}\dot{\sigma}_{ab}-2\sigma^2V_{\sigma^2}u_au_b
\label{scalar_T}
\end{eqnarray}
where $V=V(\phi,\theta,\sigma^2)$.  The terms $V_\theta$ and $V_{\sigma^2}$ are the partial derivatives of the scalar field potential with respect to $\theta$ and $\sigma^2$, respectively, and $h_{ab}\equiv g_{ab}+u_au_b$.  If there is no coupling between the aether field and the scalar field via the potential, then $V_\theta=V_{\sigma^2}=0$ and the energy momentum tensor reduces to the standard form for a minimally coupled scalar field.

\subsection{Scalar Fields having Exponential Potentials}

The simplest realisation of an inflationary theory is to consider an action with a single scalar field minimally coupled to gravity. This theory is consistent with the latest observations of cosmic microwave background temperature anisotropies data \cite{Adam:2015rua,Ade:2015xua,Ade:2015rim,Ade:2015hxq,WMAP_9year_a,WMAP_9year_b}. However,  simple quadratic and monomial potentials are coming into increasing conflict with observations since they lead to a larger tensor-to-scalar ratio than expected, resulting in renewed interest in scalar field theories having exponential potentials.
Scalar fields with exponential potentials of the form $V = V_0\e^{-2k\phi}$ are employed in alternative theories of gravity such as, for example, Brans-Dicke theory within extended inflation \cite{La:1989za,La:1989pn} and in hyper-extended inflation models \cite{Steinhardt:1990zx}.  Indeed the Ekpyrotic universe \cite{Khoury:2001wf} also employs exponential potentials. Exponential potentials appear naturally in higher dimensional frameworks, such as Kaluza-Klein and super gravity theories \cite{Cremmer:1983bf,Ellis:1983sf,Salam:1984cj,Maeda:1985es,Romans:1985tw}.  Additionally, it was shown that exponential potentials yield the least constrained form of a quintessence potential \cite{Agrawal:2018own}. In addition, exponential potentials have proven to be of interest in the study of the cosmological constraints imposed by proposed string Swampland criteria \cite{Agrawal:2018own,Akrami:2018ylq}.  From a purely mathematical perspective, the benefit in assuming an exponential potential is that the resulting dynamical system has a scaling symmetry and allows one to use dimensionless variables \cite{Coley1994a,wainwright_ellis2005,Coley:2003mj}.

Although the exponential potential for the scalar field does not lead to exponential inflation in general relativity \cite{Olive:1989nu,Linde:1987}, it does lead to power-law inflation if the potential is not too steep; that is if $k^2<\frac{1}{2}$ \cite{Kitada:1991ih,Halliwell:1986ja,Ibanez:1995zs}. Interestingly, we note that if there exists a scalar field potential that is on its own too steep to be a source of inflation, one can still recover power-law inflation by using multiple scalar fields resulting in assisted inflation \cite{Liddle:1998jc,Malik:1998gy,Copeland:1999cs,Coley:1999mj}. Here we note that in the literature the scalar field potential is sometimes written as $V(\phi)=V_0\e^{-k\phi}$, so to compare our results here and what is sometimes found in the literature one must multiply the parameter $k$ used here by $2$.


\section{The Kantowski-Sachs models coupled to a Scalar Field}

\subsection{The Geometry}

The spherically symmetric spatially homogeneous Kantowski-Sachs models under consideration have a four dimensional group of isometries acting non-transitively on three dimensional spatial hyper-surfaces \cite{wainwright_ellis2005,Stephani:2003tm}.  Here we shall assume that the aether vector field is invariant under the same group of symmetries as the metric and is consequently aligned with the symmetry adapted time coordinate and thus hyper-surface normal. In co-moving coordinates adapted to the symmetries of the metric we write $u^{a}=(1,0,0,0)$ and express the line element as
\begin{equation}
ds^2 = - dt^2 + a(t)^2 dx^2 + b(t)^2 (d\vartheta^2 + \sin^2 \vartheta  d\varphi^2).
\end{equation}

We note that the evolution equations for the more general spherically symmetric models are derived in \cite{Alhulaimi:2017} and \cite{Coley:2015qqa} using the $1+3$ formalism described in \cite{vanElst:1996dr}.  The equations here can be determined as a special case of those presented in \cite{Coley:2015qqa} with the matter source adjusted to be a scalar field.

With the given assumptions on the metric and the aether vector, the vorticity and the acceleration of the aether vector are zero, and the covariant derivative
\begin{equation}
\nabla_b u_{a}=\sigma_{ab}+\frac{1}{3}\theta(g_{ab}+u_a u_b),
\end{equation}
is determined by the expansion scalar
\begin{equation}
\theta = \nabla_au^a = \frac{\dot a}{a}+2\frac{\dot b}{b},
\end{equation}
and the shear tensor
\begin{equation}
\sigma_{ab}=u_{(a;b)}-\frac{1}{3}\theta (g_{ab}+u_a u_b),
\end{equation}
which has the form $\sigma^a_{\phantom{a}b}=\mbox{Diag}[0,-2\sigma_+,\sigma_+,\sigma_+]$ where
\begin{equation}
\sigma_+=\frac{1}{3}\left(\frac{\dot b}{b}-\frac{\dot a}{a}\right).
\end{equation}
We note that the shear scalar is
\begin{equation}
\sigma^2=\frac{1}{2}\sigma_{ab}\sigma^{ab}=3\sigma_+^2.
\end{equation}

With the definition of $T_{ab}^{\textsc{U}}$ in equation \eqref{T_ab_AE}, the effective energy density $\rho^{\textsc{U}}$, isotropic pressure, $p^{\textsc{U}}$, energy flux $q_{a}^{\phantom{a}\textsc{U}}$, and anisotropic stress $\pi_{\phantom{a}b}^{a\phantom{b}\textsc{U}}$   due to the aether field are:
\begin{eqnarray}
\rho^{\textsc{U}} & = & -\frac{1}{3}c_\theta \theta^2 -6c_\sigma \sigma_+^2,\\
p^{\textsc{U}} & = & \frac{1}{3}c_\theta \theta^2 +\frac{2}{3}c_\theta \dot\theta-6c_\sigma \sigma_+^2,\\
q_{a}^{\phantom{a}\textsc{U}}&=&0,\\
\pi_{\phantom{a}b}^{a\phantom{b}\textsc{U}}& = &2c_\sigma (\dot\sigma^a_{\phantom{a}b}+\theta\sigma^a_{\phantom{a}b}),
\end{eqnarray}
where the new parameters $c_{\theta}= (c_1 +3c_2 +c_3)$ and $c_\sigma=(c_1+c_3)$, defined before in \cite{Coley:2015qqa,Latta:2016jix,Alhulaimi:2017ocb}, allow for some efficiency in notation since the field equations are independent of any other linear combinations of the $c_i$.

The Einstein-aether field equations reduce to the following:
\begin{eqnarray}
0&=& -\frac{1}{3}(1+c_\theta)\theta^2+3(1-2c_\sigma)\sigma_+^2-{}^3R+8\pi G\rho^{\textsc{M}}, \label{Friedmann}\\
0 &=& -(1+c_\theta)\dot\theta -\frac{1}{3}(1+c_\theta) \theta^2 -6(1-2c_\sigma)\sigma_+^2 -\frac{8\pi G}{2}(\rho^{\textsc{M}}+3p^{\textsc{M}} ),\label{Raychaudhuri}\\
0&=& (1-2c_\sigma){\dot\sigma}_+ +(1-2c_\sigma)\theta\sigma_+ + \frac{1}{3b^2} -8\pi G\pi_+^{\textsc{M}},
\end{eqnarray}
where if we assume $c_\theta+1>0$, then we have the freedom to choose new units such that $\frac{8\pi G}{1+c_\theta}=1$.  We note that the curvature of the three-dimensional surfaces of homogeneity is given by
\begin{equation}
{}^3R=\frac{1}{b(t)^2}.
\end{equation}
If we define a new parameter $C = \frac{1-2c_\sigma}{1+c_\theta}$ then the explicit dependence of the field equations on the aether parameter $c_\theta$ is eliminated. However, we emphasize that $c_\theta$ is still present in the first integral \eqref{Friedmann}.  Assuming the PPN parameters for general relativity and Einstein-aether theory are the same and that there are stable positive energy modes but no vacuum \v{C}erenkov radiation, then we have the following bounds
\begin{eqnarray}
0 &\leq& c_\sigma \leq \frac{1}{2}, \nonumber \\
1-2c_\sigma &\leq& C \leq 1-\frac{3}{2}c_\sigma. \label{C_constraints}
\end{eqnarray}
on the values of $c_\sigma$ and $C$, and consequently on $c_\theta$ \cite{Alhulaimi:2017ocb}.  In particular, we note that $C\geq0$, and therefore for ease of notation in our analysis we let $C=c^2$ which inherits the following bounds $0\leq c \leq 1$ where $c=1$ is general relativity.

\subsection{The Scalar Field Potential}
We shall consider a class of generalized exponential scalar field potentials of the form
\begin{equation}
V(\phi, \theta, \sigma_+) = a_1 \e^{-2k\phi} + a_2\theta \e^{-k\phi} + a_3\sigma_+ \e^{-k\phi}\label{potential}
\end{equation}
where we shall assume that $k\geq0$ and $a_1 \geq 0$ but we put no restrictions on the signs of $a_2$ and $a_3$. In this case, from equation \eqref{scalar_T}, the effective energy density $\rho^{\textsc{M}}$, isotropic pressure $p^{\textsc{M}}$, energy flux $q_{a}^{\phantom{a}\textsc{M}}$, and anisotropic stress $\pi_{\phantom{a}b}^{a\phantom{b}\textsc{M}}$ due to the scalar field are
\begin{eqnarray}
\rho^{\textsc{M}} & = & \frac{1}{2}\dot\phi^2+ a_1 \e^{-2k\phi},\\
p^{\textsc{M}} & = & \frac{1}{2}\dot\phi^2 -  a_1 \e^{-2k\phi} - ka_2 \dot\phi \e^{-k\phi} - a_3 \sigma_+ \e^{-k\phi},\\
q_{a}^{\phantom{a}\textsc{M}} & = & 0\vphantom{\frac{1}{2}}, \\
\pi^\textsc{M}_+& = & \frac{a_3}{6}\big(\theta - k\dot\phi\big) \e^{-k\phi},
\end{eqnarray}
where $\pi_{\phantom{a}b}^{a\phantom{b}\textsc{M}}=\mbox{Diag}[0,-2\pi^\textsc{M}_+,\pi^\textsc{M}_+,\pi^\textsc{M}_+]$.
The field equation for the scalar field comes from the Klein-Gordon
equation \eqref{GeneralKleinGordoneqaution}, which becomes
\begin{equation}
0=\ddot{\phi}+\theta\dot{\phi}-2a_1k\e^{-2k\phi} -a_2k\theta \e^{-k\phi} -a_3k\sigma_+ \e^{-k\phi}.\label{KG-2}
\end{equation}

\subsection{The Dynamical System}

The evolution equations for the Einstein-aether Kantowski-Sachs model with a scalar field having a generalized exponential potential become the following system of autonomous ordinary differential equations
\begin{eqnarray}
\dot\theta &=&  -\frac{1}{3} \theta^2 -6c^2\sigma_+^2 -\psi^2 +  a_1 \e^{-2k\phi} + \frac{3}{2}ka_2 \psi \e^{-k\phi} + \frac{3}{2}a_3 \sigma_+ \e^{-k\phi},\label{DS1}\\
{\dot\sigma}_+&=&  -\theta\sigma_+ - \sigma_+^2 +\frac{1}{3c^2}\left(\frac{1}{3}\theta^2 -\frac{1}{2}\psi^2- a_1 \e^{-2k\phi}\right) +\frac{a_3}{6c^2}\big(\theta - k\psi\big) \e^{-k\phi},\\
\dot\phi&=&\psi\vphantom{\frac{1}{2}},\\
\dot\psi&=& -\theta\psi+2a_1k\e^{-2k\phi} +a_2k\theta \e^{-k\phi} +a_3k\sigma_+ \e^{-k\phi},\vphantom{\frac{1}{2}}\label{DS4}
\end{eqnarray}
with first integral
\begin{equation}
\frac{1}{1+c_\theta}\frac{1}{b^2}  = - \frac{1}{3}\theta^2+ 3c^2\sigma_+^2+\frac{1}{2}\psi^2+ a_1 \e^{-2k\phi},\label{FI}
\end{equation}
Therefore equations \eqref{DS1}-\eqref{DS4} yield a four dimensional dynamical system for the variables $(\theta,\sigma_+,\psi,\phi)$ depending on five parameters $(k,c,a_1,a_2,a_3)$ having a first integral given by equation \eqref{FI} with an additional parameter $c_\theta$.

\subsection{The Dynamical System in Normalized Variables}
We introduce normalized variables
\begin{equation}
\Phi = \frac{\sqrt{a_1}\e^{-k\phi}}{D}, \qquad
\Psi = \frac{\psi}{\sqrt{2}D},\qquad
  y = \frac{\sqrt{3}\sigma_+}{D},\qquad
  Q = \frac{\theta}{\sqrt{3}D},\label{variable1}
\end{equation}
where,since we assumed  $c_\theta+1>0$, we can define
\begin{equation}
D=\sqrt{\frac{1}{1+c_\theta}\frac{1}{b^2}+\frac{1}{3}\theta^2},\label{variable2}
\end{equation}
and a new time $\tau$ such that
\begin{equation}
\frac{d\tau}{dt}={D}.
\end{equation}
We also introduce re-scaled parameters $\tilde{a}_2=a_2/\sqrt{a_1}$ and $\tilde{a}_3=a_3/\sqrt{a_1}$.  With these variables and re-scaled parameters, the resulting system of autonomous differential equations
\begin{eqnarray}
\frac{d\Phi}{d\tau} &=&-\Phi\left(\sqrt{2}k\Psi+\chi\right),\label{dimensionlessDS1}\\
\frac{d\Psi}{d\tau} &=&-\sqrt{3}Q\Psi+k\Phi\left(\sqrt{2}\Phi+\sqrt{\frac{3}{2}}\tilde{a}_2Q+\frac{1}{\sqrt{6}}\tilde{a}_3 y\right)-\Psi\chi,\\
\frac{dy}{d\tau} &=&-\sqrt{3}Qy-\frac{1}{\sqrt{3}c^2}(1-Q^2)+\frac{\tilde{a}_3}{c^2}\Phi\left(\frac{1}{2}Q-\frac{1}{\sqrt{6}}k\Psi\right)-y\chi,\\
\frac{dQ}{d\tau} &=& \frac{1}{\sqrt{3}}(1-Q^2)(Qy-\tilde{q}),\label{dimensionlessDS4}
\end{eqnarray}
depends on only four parameters $(k,c,\tilde{a}_2,\tilde{a}_3)$.  We have defined
\begin{equation}
\chi\equiv\frac{\dot D}{D^2}=\frac{1}{\sqrt{3}}\Bigl(y(Q^2-1)-Q(\tilde{q}+1)\Bigr),
\end{equation}
and
\begin{equation}
\tilde{q}\equiv qQ^2= 2c^2y^2 +2 \Psi^2 -\Phi^2-\frac{3}{\sqrt{2}}k\tilde{a}_2\Phi\Psi-\frac{\sqrt{3}}{2}\tilde{a}_3 y\Phi,
\end{equation}
where the deceleration parameter is
\begin{equation}
q \equiv -\frac{ a \ddot{a}}{\dot{a}^2}=- \left(3\frac{\dot{\theta}}{\theta^2}+1\right). \label{def_q}
\end{equation}
The first integral \eqref{FI} becomes
\begin{equation}
f(\Phi,\Psi,y,Q)=1-c^2y^2-\Psi^2-\Phi^2=0. \label{dimensionlessconstraint}
\end{equation}

The definition of the normalized variable $\Phi$ via \eqref{variable1} implies that $\Phi\geq 0$.  Further, through the variable definitions \eqref{variable1} and \eqref{variable2} and the assumption that $c_\theta+1>0$, we have
\begin{equation}
Q^2=\frac{1}{1+\frac{3}{(1+c_\theta)}\frac{1}{b^2\theta^2}}\leq 1.
\end{equation}
Therefore we observe that the phase space is bounded and is topologically equivalent to the Cartesian product of the upper half of the unit sphere and a finite interval; i.e., $S^{2+} \times [-1,1]$.  While we have a four differential equations \eqref{dimensionlessDS1}-\eqref{dimensionlessDS4} for the variables $\Phi,\Psi,y,Q$, we also have a constraint \eqref{dimensionlessconstraint} which restricts the dynamics to three dimensions. Since the constraint equation is non-linear, it is not possible to implement the constraint globally to eliminate one of the variables.  However, it is possible to implement the constraint locally near a particular equilibrium point provided the $\nabla f$ evaluated at the equilibrium point is non-trivial.

\section{Qualitative Analysis of the Dynamical System}
\subsection{The Equilibrium Points}

The equilibria for the system \eqref{dimensionlessDS1}-\eqref{dimensionlessDS4} can be categorized depending within which invariant set $Q^2<1$, $Q=+1$, or $Q=-1$ the equilibria lies. Orbits in the $Q^2<1$ invariant set are anisotropic but have positive spatial curvature, and therefore represent proper Kantowski-Sachs type models.  We note that orbits that lie in the invariant sets $Q=+1$ and $Q=-1$ represent anisotropic spatially homogeneous models having three dimensional spatial hyper-surfaces with zero curvature, and therefore represent Bianchi type I models in general and, if in addition $y=0$, zero curvature FRW models.  Due to the fact that each point in the $Q=+1$ invariant set also has a mirror image in the $Q=-1$ invariant set, we distinguish between the two sets of equilibrium points through the use of a superscript, $''+''$ to indicate the point is in the $Q=+1$ invariant set, and $''-''$ to indicate the point is in the $Q=-1$ invariant set. All of the equilibrium points of the dynamical system \eqref{DS1}-\eqref{DS1} are summarized in Table \ref{Table1}.

\begin{table}[ht]\renewcommand{\arraystretch}{2.4}
\begin{center}
\begin{tabular}{|c|c|}
\hline
Label & Values for $(\Phi,\Psi,y,Q)$  \\
\hline
$KS_\delta$ &  $\displaystyle\left(0,0,  \frac{\delta}{c}, 2\delta c\right)$    \\
$C^{+}$ &  $\displaystyle\left(0,\cos(u), \frac{\sin(u)}{c}, 1\right)$     \\
$C^{-}$ &  $\displaystyle\left(0,-\cos(u), -\frac{\sin(u)}{c},-1\right)$     \\
$FR_\delta^+$ &  $\displaystyle\left(
\frac{-2\sqrt{3}k^2a_2+\delta\sqrt{6}\sqrt{K}}{3(2+k^2\tilde{a}_2^{\,2})},
\frac{2\sqrt{6}k+\delta\sqrt{3}k\tilde{a}_2\sqrt{K}}{3(2+k^2\tilde{a}_2^{\,2})},
0,1\right)$ \\
$FR_\delta^-$ &  $\displaystyle\left(
\frac{2\sqrt{3}k^2a_2+\delta\sqrt{6}\sqrt{K}}{3(2+k^2\tilde{a}_2^{\,2})},
\frac{-2\sqrt{6}k+\delta\sqrt{3}k\tilde{a}_2\sqrt{K}}{3(2+k^2\tilde{a}_2^{\,2})},
0,-1\right)$ \\
$BI_\delta^+$ & $\displaystyle\left(-\frac{\sqrt{3}}{2}\tilde{a}_2, \frac{\sqrt{6}}{2k},\frac{\delta}{2kc}\sqrt{-K},1\right)$  \\
$BI_\delta^-$ & $\displaystyle\left(\frac{\sqrt{3}}{2}\tilde{a}_2, -\frac{\sqrt{6}}{2k},-\frac{\delta}{2kc}\sqrt{-K},-1\right)$  \\
\hline
\end{tabular}
\end{center}
\caption{Summary of all equilibrium points.  The non-isolated circles of equilibrium points, $C^+$ and $C^-$, are parameterized by $u\in(-\pi,\pi]$ where $K=6+3k^2\tilde{a}_2^{\,2}-4k^2$ and $\delta=\pm 1$. The superscript indicates whether the point is in the $Q=+1$ or $Q=-1$ invariant set, with no superscript indicating the point is in neither. }\label{Table1}
\end{table}

\subsubsection{Equilibrium points: \texorpdfstring{$Q^2<1$}{Qless1} -- Kantowski-Sachs}

\paragraph{Vacuum Kantowski-Sachs Equilibrium Points}

The only equilibrium points in the four dimensional invariant set $Q^2<1$ consist of the pair of points
$$KS_\delta=(0,0,  \frac{\delta}{c}, 2\delta c) \quad \mbox{where\ } \delta = \pm1,$$
which represent Kantowski-Sachs anisotropic vacuum solutions with positive spatial curvature and positive deceleration parameter. $KS_+$ represent expanding models while $KS_-$ represent contracting models.
These points are in the phase space only if $c<\frac{1}{2}$, and therefore their existence is directly attributable to the existence of the aether field since in general relativity $c=1$ and these points do not exist.  We note the following limits:
$$
\lim_{c\to\frac{1}{2}^-}\left(KS_{+}\right) = C^+ \mbox{\qquad and \qquad} \lim_{c\to\frac{1}{2}^-}\left(KS_{-}\right) = C^-.
$$

Before determining the local stability of these points, we first calculate
$$\nabla f\big{|}_{KS_\delta}=[0,0,-2c\delta,0],$$
which indicates that one can eliminate the variable $y$ locally near the equilibrium points via the substitution $y=\frac{\delta}{c}\sqrt{1-\Phi^2-\Psi^2}$.
The eigenvalues for the resulting three dimensional dynamical system at the points $KS_{\delta}$ are
$$\frac{\sqrt{3}\delta(1+2c^2)}{3c}, \frac{\sqrt{3}\delta(1-4c^2)}{3c},\frac{\sqrt{3}\delta(1-4c^2)}{3c}.$$

Therefore, the point $KS_{+}$ is a local source when it is in the physical phase space; i.e., $c< \frac{1}{2}$. Further, the point $KS_{-}$ is a non-inflationary local sink when it is in the physical phase space; i.e., $c < \frac{1}{2}$.

\subsubsection{Equilibrium points: \texorpdfstring{$Q=+1$}{Qis1}  -- Expanding Bianchi I}

\paragraph{Kasner-like Equilibrium Points}

In the $Q=+1$ invariant set, there exists a circle of non-isolated equilibria given by
$$C^{+}=(0,\cos(u), c^{-1}\sin(u), 1) \mbox{\ with\ } u\in(-\pi,\pi],$$
which lies on the boundary of the physical space space.

To determine the local stability, we first calculate
$$\nabla f\big{|}_{C^{+}}=[0,-2\cos(u),-2\sin(u),0] $$
which indicates that one can eliminate the variable $y$ locally near the equilibrium point via the substitution $y=\frac{1}{c}\sqrt{1-\Phi^2-\Psi^2}$ for $y\geq0$ or $y=-\frac{1}{c}\sqrt{1-\Phi^2-\Psi^2}$ for $y<0$.

The eigenvalues for the resulting three dimensional system at the point $C^{+}$ are
$$0,\sqrt{3}-\sqrt{2}k\cos(u),\frac{2\sqrt{3}(2c-|\sin(u)|)}{3c}.$$
Due to the non-isolated nature of the equilibria, one of the eigenvalues is zero.  We also note that the eigendirections associated with the first two eigenvalues spans the two-dimensional $Q=+1$ invariant set.

In the $Q=+1$ invariant set, the local stability of $C^+$ is independent of the aether parameter $c$.  If $k^2<\frac{3}{2}$, then each point of $C^+$ acts like a source.  If $k^2>\frac{3}{2}$, then most of $C^+$ is a source, but for values of $\Phi>\frac{\sqrt{3}}{\sqrt{2}k}$ the points are sinks.

\begin{figure}[h]
$\begin{array}{rl}
\includegraphics[width=0.5\textwidth]{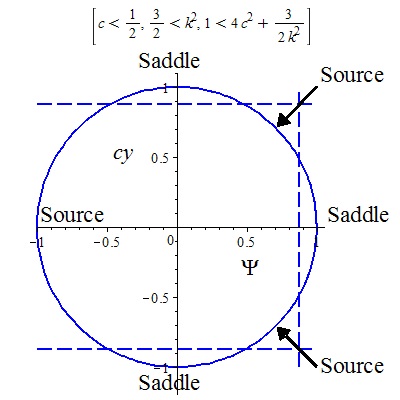}  &
\includegraphics[width=0.5\textwidth]{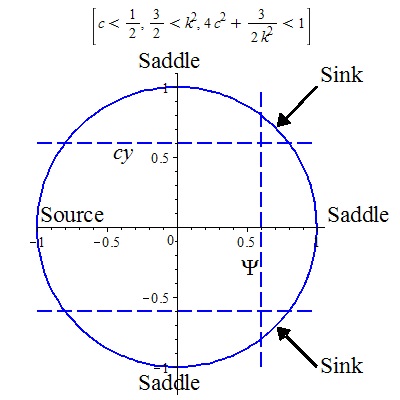}
\end{array}$
\caption{The figures indicate the local stability in the full three dimensional phase space of the non-isolated line of equilibria $C^+$ for the range of parameter values listed above. }\label{C_stability}
\end{figure}

In the full three-dimensional phase space the local stability of the non-isolated line of equilibria is summarized as:
\begin{itemize}
\item If $k^2<\frac{3}{2}$ then
    \begin{itemize}
    \item if $c>\frac{1}{2}$ then the entire $C^{+}$ is a source;
    \item if $c<\frac{1}{2}$ then most of $C^+$ is a source, but two parts of $C^{+}$ have a one dimensional stable manifold and are therefore saddles;
    \end{itemize}
\item if $k^2>\frac{3}{2}$ then
    \begin{itemize}
    \item if $c>\frac{1}{2}$  most of $C^+$ is a source, but part of $C^+$ has a one dimensional stable manifold in the $Q=+1$ set and therefore the point is a saddle;
    \item if $c<\frac{1}{2}$  but $4c^2+\frac{3}{2k^2}>1$ then a part of $C^{+}$ is a source and there are three dis-connected parts of $C^{+}$ with one dimensional stable manifolds (saddles), separated by sources (See Figure \ref{C_stability});
    \item if $c<\frac{1}{2}$  but $4c^2+\frac{3}{2k^2}<1$, then a part of $C^{+}$ is a source and there are two sections of $C^{+}$ with a two dimensional stable manifold (sink) which is surrounded on either side by saddles (See Figure \ref{C_stability});
    \end{itemize}
\end{itemize}

\paragraph{Isotropic Equilibrium Points}

The two points
$$
FR^+_{\delta} = (\Phi_{eq},\Psi_{eq},0,1)
=\left(
\frac{-2\sqrt{3}k^2{\tilde a}_2+\delta\sqrt{6}\sqrt{K}}{3(2+k^2\tilde{a}_2^{\,2})},
\frac{2\sqrt{6}k+\delta\sqrt{3}k\tilde{a}_2\sqrt{K}}{3(2+k^2\tilde{a}_2^{\,2})},
0,1\right)
$$
represent zero curvature expanding isotropic models (i.e, FRW) with a scalar field where $\delta=\pm1$. For ease of presentation we have defined the parameter
$$K = 6+3{\tilde a}_2^2k^2-4k^2=4k^2\left(\frac{3}{2k^2}+\frac{3{\tilde a}_2^2}{4}-1\right).$$
The point $FR^+_-$ exists in the physical phase space if
$$k^2\geq \frac{3}{2} \mbox{\qquad and\qquad} {\tilde a}_2<0,\  \frac{3}{2k^2}+\frac{3{\tilde a}_2^2}{4}>1 .$$
On the other hand, the point $FR^+_+$ exists in the physical phase space if
\begin{eqnarray*}
&& k^2\leq \frac{3}{2} \mbox{\qquad  or }\\
&& k^2\geq \frac{3}{2} \mbox{\qquad  and\qquad }{\tilde a}_2<0,\  \frac{3}{2k^2}+\frac{3{\tilde a}_2^2}{4}>1 .
\end{eqnarray*}
It is interesting to note that the existence of the aether/scalar field coupling in the scalar field potential allows these isotropic points to exist for larger values of $k$ than would be permitted in general relativity.  We note the following limits on the equilibrium points
$$\lim_{K  \to 0^+} \left(FR^+_{\delta}\right) = BI^+_{\delta} \mbox{\qquad and \qquad }
\lim_{k^2\to\frac{3}{2}^+} \left(FR^+_{-}\right)=
\lim_{k^2\to\frac{3}{2}^-} \left(FR^+_{+}\right) = C^{+}.$$
Additionally, these equilibrium points, $FR^+_\delta$, represent power law inflationary cosmological models if
\begin{equation}
\tilde{q}\big{|}_{FR^+_{\delta}}=\sqrt{6}k\left(\Psi_{eq}-\frac{1}{\sqrt{6}k}\right)<0.\label{q_FRW}
\end{equation}

To determine the local stability, we first calculate
$$\nabla f\big{|}_{FR^{+}_\delta}=[-2\Phi_{eq},-2\Psi_{eq},0,0], $$
which indicates that we are able to eliminate the variable $\Phi$ locally near the equilibrium point via \eqref{dimensionlessconstraint}.
The eigenvalues for the points ${FR^+_{\delta}}$ are
\begin{eqnarray*}
&&\frac{\sqrt{3}}{3(2+k^2{\tilde a}_2^2)}\left(-K  + \mbox{sgn}(\tilde{a}_2)\delta\sqrt{K^2+(2+k^2{\tilde a}_2^2)(2k^2-3)K}\right)\quad [\times 2], \\
&&\frac{2\sqrt{3}}{3(2+k^2{\tilde a}_2^2)}\left((4k^2-2-k^2{\tilde a}_2^2) + \mbox{sgn}(\tilde{a}_2)\delta\sqrt{2\tilde{a}_2^2k^4K }\right).
\end{eqnarray*}
These eigenvalues can also be expressed nicely as
$$\sqrt{2}k\left(\Psi_{eq}-\frac{3}{\sqrt{6}k}\right) \quad [\times 2], \qquad
 2\sqrt{2}k\left(\Psi_{eq}-\frac{1}{\sqrt{6}k}\right).$$
Interestingly enough, this pair of repeated eigenvalues correspond to the eigendirections spanning the $Q=+1$ invariant set.
Using \eqref{q_FRW} we conclude that; if $\Psi_{eq}<\frac{1}{\sqrt{6}k}$, then the equilibrium point $FR^{+}_\delta$ is an inflationary sink; if $\frac{1}{\sqrt{6}k}<\Psi_{eq}<\frac{3}{\sqrt{6}k}$, then the equilibrium point $FR^{+}_\delta$ is a non-inflationary saddle with a two dimensional stable manifold; and if $\frac{3}{\sqrt{6}k}<\Psi_{eq}$, then the equilibrium point $FR^{+}_\delta$ is a non-inflationary source.

It can be shown that for the equilibrium point $FR^+_-$, $\Psi_{eq}>\frac{3}{\sqrt{6}k}$ for all parameter values for which the point exists.  Therefore, the equilibrium point $FR^+_-$ is unstable and represents a non-inflationary source.

\begin{figure}[h]
\begin{center}
\includegraphics[width=0.5\textwidth]{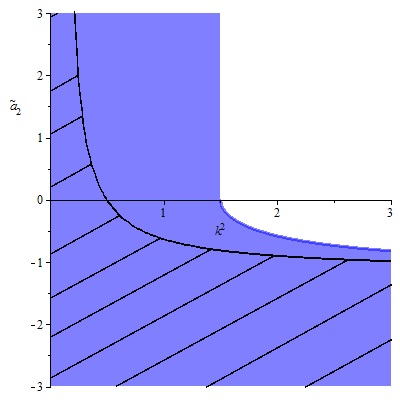}
\end{center}
\caption{The shaded part of the figure indicates the portion of the parameter space $(k^2,\tilde{a}_2)$ in which the point $FR^+_{+}$  exists in the physical phase space. The hash marks on the graph indicate parameter values in which equation \eqref{inflation1_condition} is satisfied and the point $FR^+_{+}$ is consequently a stable inflationary attractor. Note the existence of large values of $k$ for which this inflationary attractor exists.}\label{P3_Existence_Inflation}
\end{figure}

For the equilibrium point $FR^+_+$, it can be shown that $\Psi_{eq}<\frac{3}{\sqrt{6}k}$ for all parameter values for which the point exists. Therefore, the equilibrium point $FR^+_+$ has at a minimum a two dimensional stable manifold which lies in the $Q=+1$ invariant set.   Further, if the point is inflationary then the third eigenvalue is negative and the point is stable is the full phase space. The point $FR^+_{+}$ is an inflationary sink if
\begin{equation}
k^2\leq\frac{1}{6}, \mbox{\qquad\qquad or\qquad\qquad } k^2>\frac{1}{6}, \ \tilde{a}_2<\frac{\sqrt{2}(1-2k^2)}{k\sqrt{6k^2-1}}\label{inflation1_condition}
\end{equation}
See Figure \ref{P3_Existence_Inflation} which illustrates the existence and inflationary criteria for $FR^+_{+}$.

\paragraph{Bianchi type I Points}

The two points
$$BI^+_{\delta}=\left(-\frac{\sqrt{3}}{2}\tilde{a}_2, \frac{\sqrt{6}}{2k},\frac{\delta}{2kc}\sqrt{-K},1\right)$$
represent anisotropic Bianchi type I models with a scalar field, where $\delta=\pm1$. Since $q=2$ at these equilibrium points they are not inflationary.  These equilibrium points exist in the physical phase space if $K<0$ and $\tilde{a}_2<0$.
We note that
$$\lim_{K  \to 0^-} \left(BI^+_{\delta}\right) = FR^+_{\delta} \mbox{\qquad and \qquad } \lim_{\tilde{a}_2\to 0^-} \left(BI^+_{\delta}\right) = C^{+}.$$

Since
$$\nabla f\big{|}_{BI^{+}_\delta}=\left[\sqrt{3}\tilde{a}_2,-2\frac{\sqrt{6}}{k},-\frac{\delta c\sqrt{-K}}{k},0\right] $$
we can eliminate the variable $y$ locally near each of the equilibrium points.  The eigenvalues for the resulting three dimensional system in the variables $\Phi,\Psi, Q$ at the points $BI^{+}_\delta$ are
$$\frac{\sqrt{3}}{3kc}\left(4kc-\delta\sqrt{-K}\right),\pm\frac{\sqrt{6}\tilde{a}_2\sqrt{-K}}{4}i.$$
The first eigenvalue is positive at the point $BI^+_{-}$ while the sign of the first eigenvalue for $BI^+_{+}$ depends on the sign of $K+(4kc)^2$.

The second and third eigenvalues for both $BI^+_{-}$ and $BI^+_{+}$ are purely imaginary and additional analysis is required to determine the local stability.  Recognizing that the eigenspace spanned by the purely imaginary eigenvalues (or center manifold) is the invariant set $Q=+1$, we can restrict our analysis of this point to this two dimensional invariant set in $\Phi$ and $\Psi$.  Translating the origin to the point $\left(-\frac{\sqrt{3}}{2}\tilde{a}_2,\frac{\sqrt{6}}{2k}\right)$, and applying a rotation to obtain the Jordan canonical form \cite{hale1996dynamics} and then changing to polar coordinates we define
\begin{eqnarray}
r \cos(\theta)&=& \frac{1}{\sqrt{2}k(3\tilde{a}_2^2-4)}\left(\Phi+\frac{\sqrt{3}}{2}\tilde{a}_2\right),\\
r \sin(\theta) &=& \frac{3\tilde{a}_2}{2k\sqrt{-K}(3\tilde{a}_2^2-4)}\left(\Phi+\frac{\sqrt{3}}{2}\tilde{a}_2\right) + \frac{1}{2\sqrt{2}\sqrt{-K}}\left(\Psi-\frac{\sqrt{6}}{2k}\right).
\end{eqnarray}
The resulting system of differential equations for $r$ and $\theta$ becomes
\begin{eqnarray}
r' &=&-\frac{kr^2}{\sqrt{-K}}\left( K\tilde{a}_{2}\sin\theta-2\sqrt{2}\sqrt{-K}\cos\theta\right)\left(\left[2\sqrt{6}k(3\tilde{a}_2^2-4)\cos\theta\right] r  +3\tilde{a}_2\right), \label{r_prime}\\
\theta' &=& -\frac{\sqrt{-K}}{2\sqrt{6}}\left(\left[2\sqrt{6}k(3\tilde{a}_2^2-4)\cos\theta\right]r  +3\tilde{a}_2\right).\label{theta_prime}
\end{eqnarray}
Notice that the righthand sides of both \eqref{r_prime} and \eqref{theta_prime} contain the factor $$\left[2\sqrt{6}k(3\tilde{a}_2^2-4)\cos\theta\right]r  +3\tilde{a}_2$$ which is non-zero in a neighborhood of the origin and can be cancelled in $\frac{dr}{d\theta}$.  It is straightforward to construct a first integral since the differential equation  $\frac{dr}{d\theta}$ is separable;  the solution is
\begin{equation}
r(\theta)=\left[\frac{1}{r_0}-2\sqrt{6}k\tilde{a}_2(\cos\theta-1)+\frac{8\sqrt{3}k}{\sqrt{-K}}\sin\theta\right]^{-1},
\end{equation}
where $r_0=r(0)$ is the constant of integration.  The solution clearly shows an infinite family of periodic orbits for small values of $r_0$. Since $\tilde{a}_2<0$, from \eqref{theta_prime} we observe that these periodic orbits tend to revolve around the point in a counter-clockwise fashion. As one increases the value of $r_0$, these solution curves will eventually intersect either the line of equilibria $C^+$ or the points $FR^+_{\delta}$ on the boundary and will no longer represent periodic orbits. Therefore, the points $BI^+_{\delta}$ are non-linear centers when they exist in the $Q=1$ invariant set.

In summary, the point $BI^+_{-}$ is always an asymptotically unstable non-linear center in the full phase space. Similarly, the point $BI^+_{+}$ is also an asymptotically unstable non-linear center in the full phase space if $K+(4kc)^2>0$, or equivalently, if
$$4c^2+\frac{3}{2k^2}+\frac{3\tilde{a}_2^2}{4}> 1.$$
However, if
$$4c^2+\frac{3}{2k^2}+\frac{3\tilde{a}_2^2}{4}< 1,$$ then the point $BI^+_{+}$ is an asymptotically stable non-linear center and thus there exists asymptotically stable periodic orbits in a local neighborhood of $BI^+_{+}$ in the full phase space.  It is very interesting to note that the existence of these periodic orbits is a direct result of the aether/scalar field coupling.

\subsubsection{Equilibrium points: \texorpdfstring{$Q=-1$}{Qisneg1} -- Contracting Bianchi I}

We note that the dynamical system \eqref{dimensionlessDS1}-\eqref{dimensionlessDS4} is invariant under the transformation
\begin{equation}
(\tau,\Phi,\Psi,y,Q,{\tilde{a}_2})\to (-\tau,\Phi,-\Psi,-y,-Q,-{\tilde{a}_2}), \label{mapping}
\end{equation}
and therefore the equilibrium points in the $Q=+1$ invariant set are mapped to the equilibrium points in the $Q=-1$ invariant set via;
\begin{equation}
C^+\to C^-,\qquad FR^+_\delta\to FR^-_\delta, \qquad  BI^+_\delta\to BI^-_\delta.
\end{equation}

Further, due to this symmetry, if an equilibrium point in the $Q=1$ invariant set is asymptotically stable for $\alpha<{\tilde a}_2<\beta$, then it's corresponding point in the $Q=-1$ invariant set is asymptotically unstable for $\alpha<-{\tilde a}_2<\beta$. That is, the unstable and stable manifolds for each equilibrium point in the $Q=1$ set for $\alpha<{\tilde a}_2<\beta$ become the stable and unstable manifolds, respectively, for the corresponding equilibrium point in the $Q=-1$ set with $\alpha<-{\tilde a}_2<\beta$. Simply put, the eigenvalues for each point in the $Q=-1$ invariant set are the negative of the eigenvalues (with ${\tilde a}_2 \to -{\tilde a}_2$) for the corresponding equilibrium points in the $Q=1$ invariant set. We note that the normalized deceleration parameter behaves like $\tilde q \to \tilde q$ under equation \eqref{mapping} indicating that the inflationary behaviour of points in the $Q=-1$ set is the same as their corresponding points in the $Q=+1$ invariant set. We summarize the local stability below without explicitly showing the calculations.

\paragraph{Kasner-like Equilibrium Points}

The non-isolated line of equilibria $C^{-}$ exists for all values of the parameters. Its local stability is:
\begin{itemize}
\item if $k^2<\frac{3}{2}$ then
    \begin{itemize}
        \item if $c>\frac{1}{2}$ then the entire $C^{-}$ is an attractor in the full phase space;
        \item if $c<\frac{1}{2}$ then part of $C^{-}$ is an attractor but there are two parts of $C^{-}$ which only have a one dimensional stable manifold and are therefore saddles;
    \end{itemize}
\item if $k^2>\frac{3}{2}$ then
    \begin{itemize}
        \item if $c>\frac{1}{2}$  then a part of $C^{-}$ is an attractor and there are two parts of $C^-$ which only have a one dimensional stable manifold and are therefore saddles;
    \item if $c<\frac{1}{2}$ but $4c^2+\frac{3}{2k^2}>1$ then a part of $C^{-}$ is an attractor and there are three dis-connected parts of $C^{-}$ with one dimensional stable manifolds (saddles), separated by attractors; 
    \item if $c<\frac{1}{2}$ but $4c^2+\frac{3}{2k^2}<1$, then a part of $C^{-}$ is an attractor and there are two sections of $C^{-}$ with a two dimensional unstable manifold (source) which is surrounded on either side by saddles;
    \end{itemize}
\end{itemize}

\paragraph{Isotropic Equilibrium Points}

The point $FR^-_-$ exists in the physical phase space if
$$k^2\geq \frac{3}{2} \mbox{\qquad and\qquad} {\tilde a}_2>0,\  \frac{3}{2k^2}+\frac{3{\tilde a}_2^2}{4}>1 .$$
On the other hand, the point $FR^-_+$ exists in the physical phase space if
\begin{eqnarray*}
&& k^2\leq \frac{3}{2}  \mbox{\qquad  or }\\
&& k^2\geq \frac{3}{2} \mbox{\qquad  and\qquad }{\tilde a}_2>0,\  \frac{3}{2k^2}+\frac{3{\tilde a}_2^2}{4}>1 .
\end{eqnarray*}
The equilibrium point $FR^-_-$ is asymptotically stable when it exists and represents a non-inflationary attractor. Further, the point $FR^-_{+}$ is unstable but inflationary in the $Q=-1$ invariant set.  In the full phase space, the point $FR^-_{+}$ is an inflationary source if
\begin{equation}
k^2\leq\frac{1}{6}, \mbox{\qquad\qquad or\qquad\qquad } k^2>\frac{1}{6}, \ \tilde{a}_2>-\frac{\sqrt{2}(1-2k^2)}{k\sqrt{6k^2-1}}\label{inflation2_condition}
\end{equation}
otherwise it is a saddle with a one-dimensional stable manifold.

\paragraph{Bianchi type I Points}

The equilibrium points $BI^-_{\delta}$ exist in the physical phase space if $K<0$ and $\tilde{a}_2>0$. The point $BI^-_{-}$ is always an asymptotically stable non-linear center. Similarly, the point $BI^-_{+}$ is also an asymptotically stable non-linear center in the full phase space if
$$4c^2+\frac{3}{2k^2}+\frac{3\tilde{a}_2^2}{4}> 1.$$
However, if
$$4c^2+\frac{3}{2k^2}+\frac{3\tilde{a}_2^2}{4}< 1$$ then the point $BI^-_{+}$ becomes an asymptotically unstable non-linear center.

\subsection{Invariant Sets}

\subsubsection{The \texorpdfstring{$Q=+1$}{Qis1} Invariant Set}

The set $Q=1$ is a lower dimensional invariant set of the dynamical system. Orbits in this set represent expanding Bianchi type I Einstein-aether models containing a scalar field having an exponential potential with a non-trivial coupling between the aether and scalar fields.  Without loss of generality, the variable $y^2$ can be globally replaced in the dynamical system \eqref{dimensionlessDS1}-\eqref{dimensionlessDS4} to obtain a two dimensional system in the variables $\Phi$ and $\Psi$.  The equilibrium points in this set are $BI^+_\delta$, $FR^+_\delta$ and $C^+$.  The dynamics in this set depend upon the values of $k$ and the coupling parameter ${\tilde a}_2$, but interestingly enough are independent of the aether parameter $c$. We note that none of these points are saddle points in the invariant set.

\begin{figure}[p]
$$\begin{array}{rl}
\includegraphics[width=0.4\textwidth]{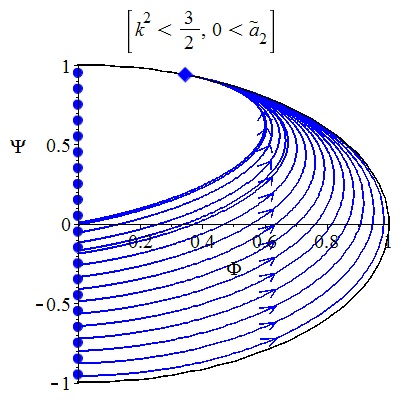}  &
\includegraphics[width=0.4\textwidth]{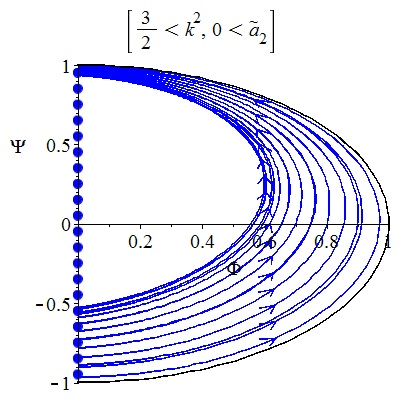}  \\
\includegraphics[width=0.4\textwidth]{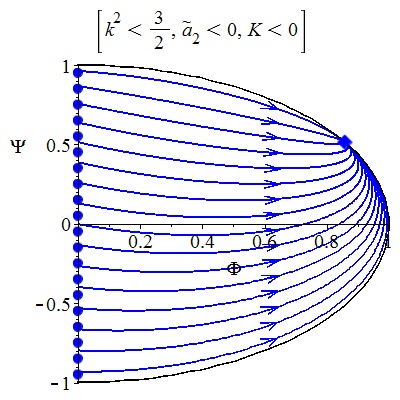}  &
\includegraphics[width=0.4\textwidth]{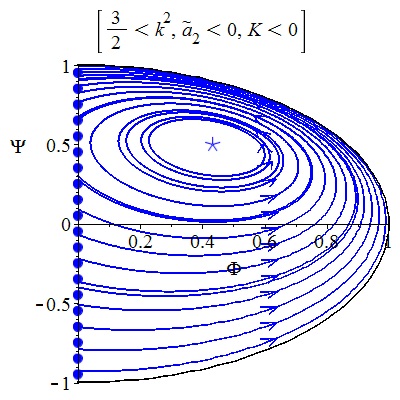}  \\
\includegraphics[width=0.4\textwidth]{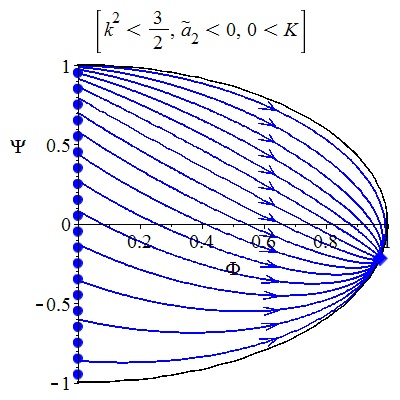}  &
\includegraphics[width=0.4\textwidth]{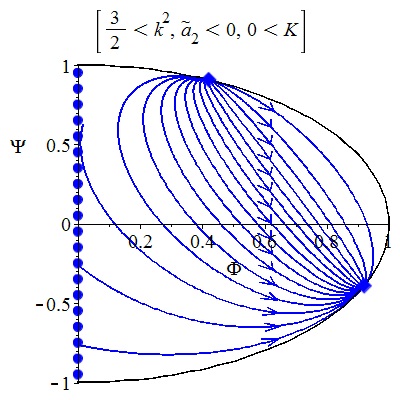}
\end{array}$$
\caption{Phase portraits in the $Q=1$ invariant set. The solid circles represent the line of equilibrium points $C^{+}$, while an asterisk represents the points $BI_\delta^+$ and diamonds represent the points $FR_\delta^+$.} \label{Phase_Portraits1}
\end{figure}

If $k^2<\frac{3}{2}$, then $FR^+_+$ is asymptotically stable while $C^+$ is asymptotically unstable. If $k^2>\frac{3}{2}$, then a part of $C^+$ is asymptotically stable and a part is asymptotically unstable. In addition, if $k^2>\frac{3}{2}$, ${\tilde a}_2<0$ and $K>0$, then $FR^+_+$ is asymptotically stable while $FR^+_-$ is asymptotically unstable. Further, we emphasize the existence of an infinite number of periodic orbits if $k^2>\frac{3}{2}$, $\tilde{a}_2 < 0$ and $K < 0$.  A few phase portraits are provided to illustrate the different asymptotically special behaviours in this lower dimensional set (See Figure \ref{Phase_Portraits1}).

\subsubsection{The \texorpdfstring{$Q=-1$}{Qisneg1} Invariant Set}

The set $Q=-1$ is a lower dimensional invariant set of the dynamical system. Orbits in this set represent contracting Bianchi type I Einstein-aether models containing a scalar field having an exponential potential with a non-trivial coupling between the aether and scalar fields.  Without loss of generality, the variable $y^2$ can be globally replaced in the dynamical system \eqref{dimensionlessDS1}-\eqref{dimensionlessDS4} to obtain a two dimensional system in the variables $\Phi$ and $\Psi$.  The equilibrium points in this set are $BI^-_\delta$, $FR^-_\delta$ and $C^-$.  The dynamics in this set depends upon the values of $k$ and the coupling parameter ${\tilde a}_2$, but are independent of the aether parameter $c$. We note that none of these points are saddle points in the invariant set.

\begin{figure}[p]
$$\begin{array}{rl}
\includegraphics[width=0.4\textwidth]{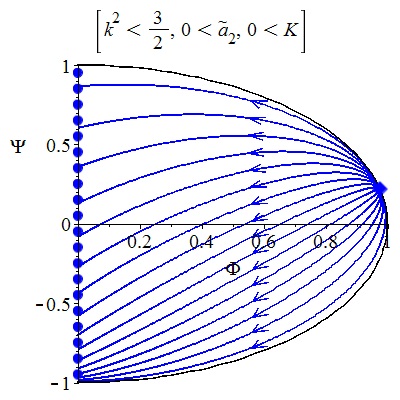}  &
\includegraphics[width=0.4\textwidth]{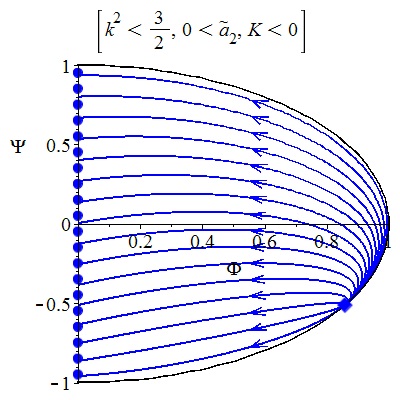}  \\
\includegraphics[width=0.4\textwidth]{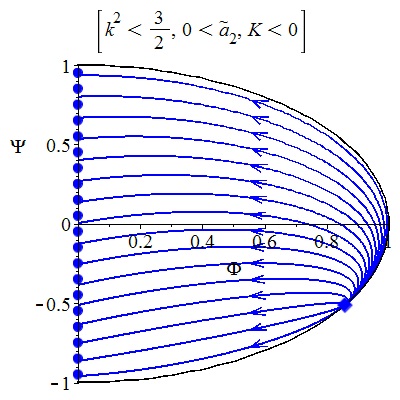}  &
\includegraphics[width=0.4\textwidth]{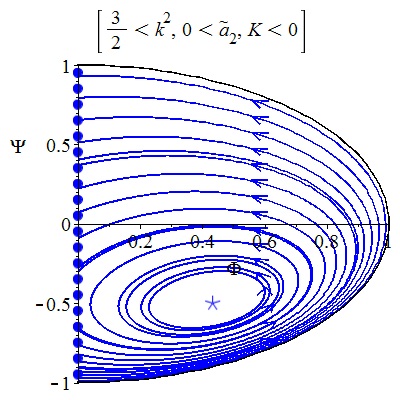}  \\
\includegraphics[width=0.4\textwidth]{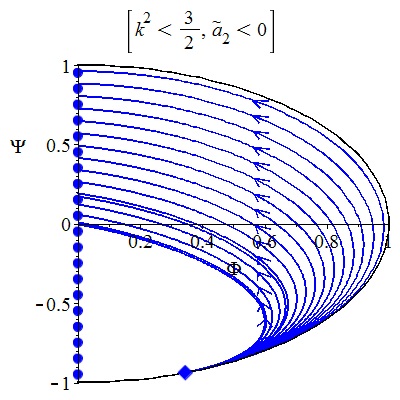}  &
\includegraphics[width=0.4\textwidth]{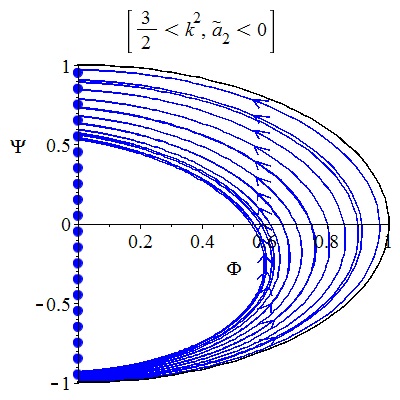}
\end{array}$$
\caption{Phase portraits in the $Q=-1$ invariant set.  The solid circles represent the line of equilibrium points $C^{-}$, while an asterisk represents the points $BI_\delta^-$ and diamonds represent the points $FR_\delta^-$.  Note how the figures are reflections of the figures in Figure \ref{Phase_Portraits1} with the direction of flow reversed.} \label{Phase_Portraits2}
\end{figure}

If $k^2<\frac{3}{2}$, then $FR^-_+$ is asymptotically unstable while $C^-$ is asymptotically stable.  If $k^2>\frac{3}{2}$, then a part of $C^-$ is asymptotically unstable and a part is asymptotically stable. In addition, if $k^2>\frac{3}{2}$, ${\tilde a}_2>0$ and $K>0$, then $FR^-_+$ is asymptotically unstable while $FR^-_-$ is asymptotically stable. Further, we emphasize again the existence of an infinite number of periodic orbits if $k^2>\frac{3}{2}$, $\tilde{a}_2 >0$ and $K < 0$.  A few phase portraits are provided to illustrate the different asymptotically special behaviours in this lower dimensional set (See Figure \ref{Phase_Portraits2}).

\subsubsection{The \texorpdfstring{$\Phi=0$}{Phizero} Invariant Set}

\begin{figure}[p]
$$\begin{array}{rl}
\includegraphics[width=0.4\textwidth]{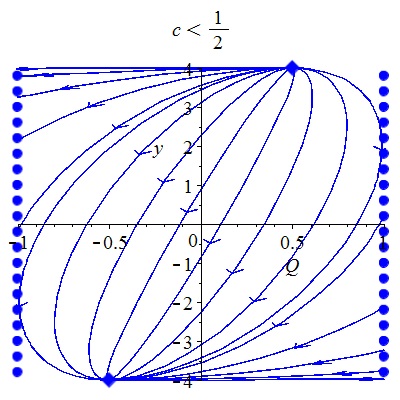}  &
\includegraphics[width=0.4\textwidth]{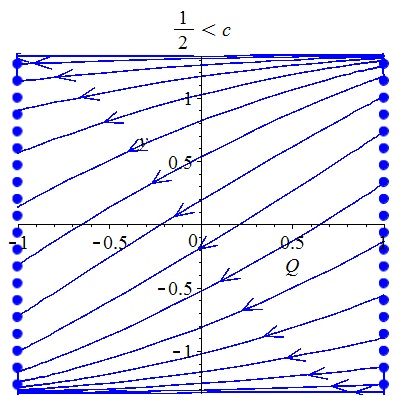}
\end{array}$$
\caption{Phase portraits in the two dimensional invariant set $\Phi=0$ for the range of parameter values listed above. The solid circles represent the two lines of equilibrium points $C^{-}$ (left-hand side of each) and $C^{+}$ (right-hand side of each), while a diamond represents the points $KS_\delta$. }\label{Phase_Portraits3}
\end{figure}

The set $\Phi=0$ is another physically interesting invariant set of the dynamical system.  This set represents Kantowski-Sachs Einstein-aether models containing a massless scalar field. Without loss of generality, the variable $\Psi^2$ can be globally replaced in the dynamical system \eqref{dimensionlessDS1}-\eqref{dimensionlessDS4} to obtain a two dimensional system in the variables $Q$ and $y$.  The equilibrium points in this set are $KS_\delta$ and $C^\delta$.  The dynamics in this set only depend on the value of the aether parameter $c$. It is straightforward to conclude that if $c<\frac{1}{2}$, then the point $KS_+$ is asymptotically unstable while the point $KS_-$ is asymptotically stable, while parts of both $C^+$ and $C^-$ are asymptotically stable and asymptotically unstable. For $c\geq \frac{1}{2}$, we see that the entire $C^+$ is asymptotically unstable while all of $C^-$ is asymptotically stable.  We include two typical phase portraits (See Figure \ref{Phase_Portraits3}).


\section{Discussion}

\subsection{Summary}
\begin{figure}[p]
$$\includegraphics[width=0.9\textwidth]{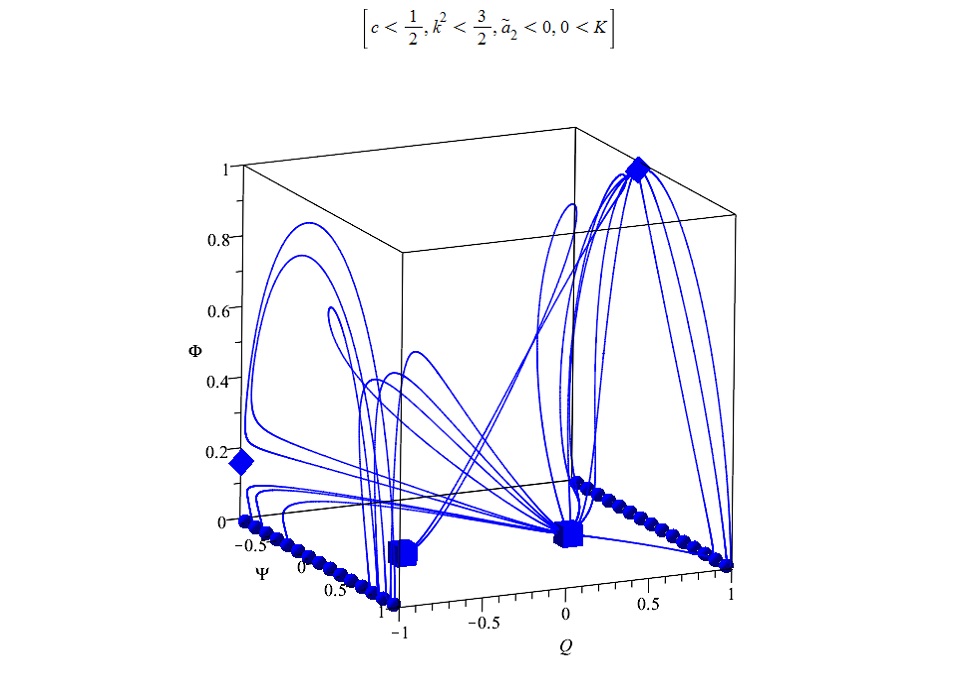}$$
$$\includegraphics[width=0.9\textwidth]{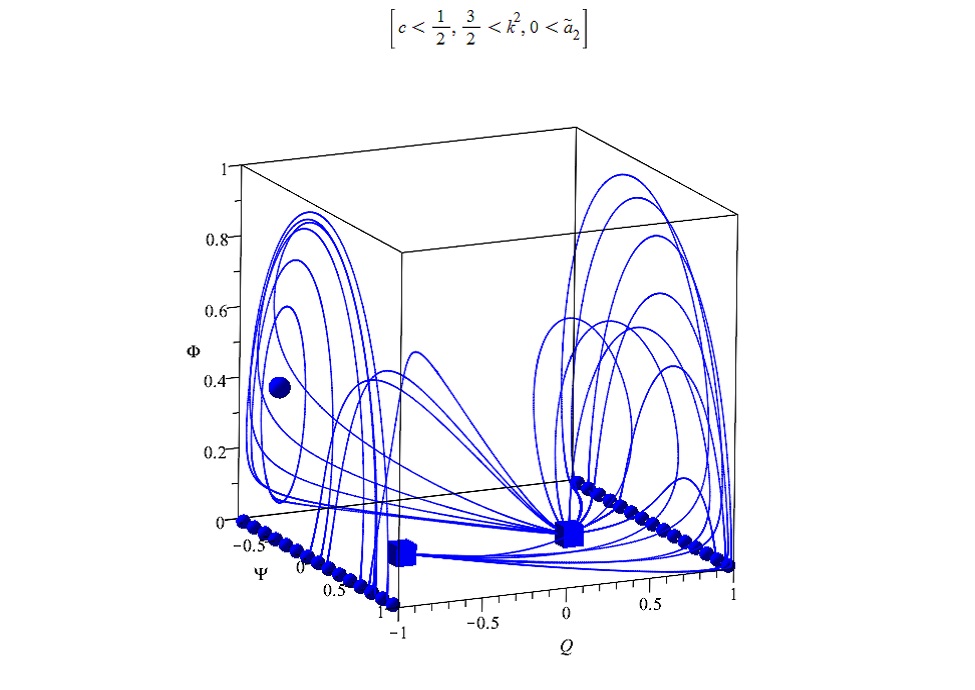}$$
\caption{A selection of phase portraits of the Kantowski-Sachs models in the full phase space.  The small circles represent the line of equilibrium points $C^+$ and $C^{-}$, while the large sphere represents the points $BI_\delta^+$ and $BI_\delta^-$ and diamonds represent the points $FR_\delta^+$ and $FR_\delta^-$.  Note the existence of the two $KS_\delta$ points (represented by boxes) when $c<\frac{1}{2}$.} \label{Phase_Portraits3D1}
\end{figure}

\begin{figure}[p]
$$\includegraphics[width=0.9\textwidth]{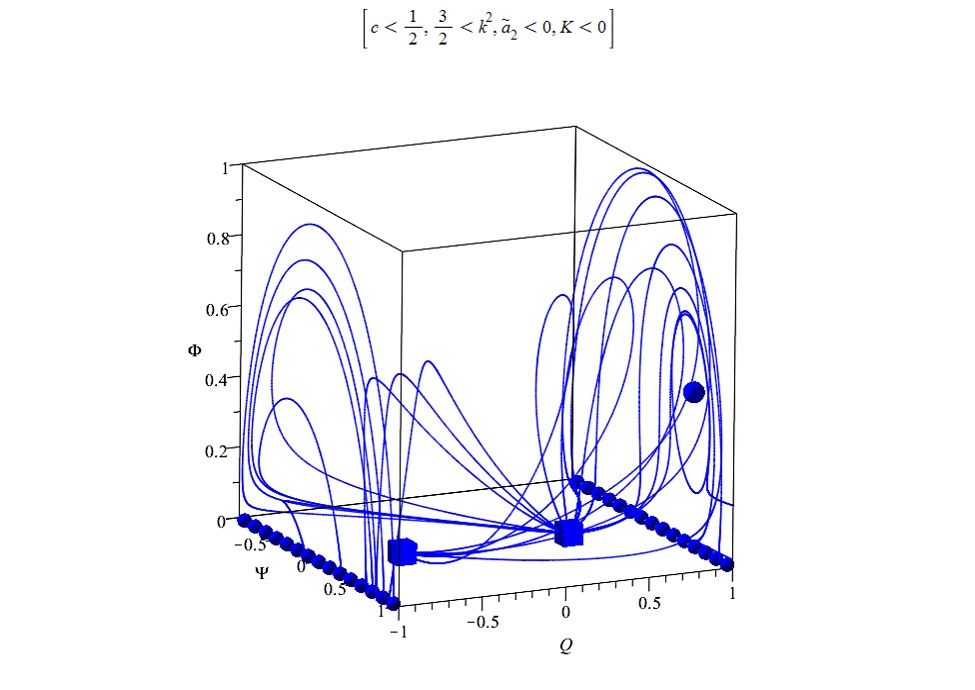}$$
$$\includegraphics[width=0.9\textwidth]{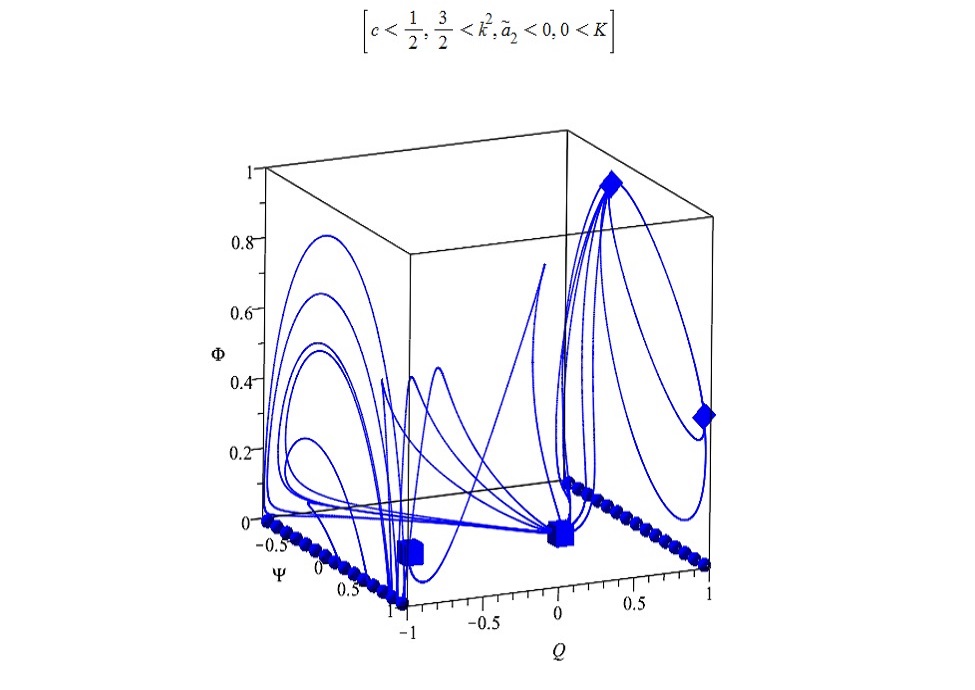}$$
\caption{A selection of phase portraits of the Kantowski-Sachs models in the full phase space.  The small circles represent the line of equilibrium points $C^+$ and $C^{-}$, while the large sphere represents the points $BI_\delta^+$ and $BI_\delta^-$ and diamonds represent the points $FR_\delta^+$ and $FR_\delta^-$.  Note the existence of the two $KS_\delta$ points (represented by boxes) when $c<\frac{1}{2}$.} \label{Phase_Portraits3D2}
\end{figure}

\begin{figure}[p]
$$\includegraphics[width=0.9\textwidth]{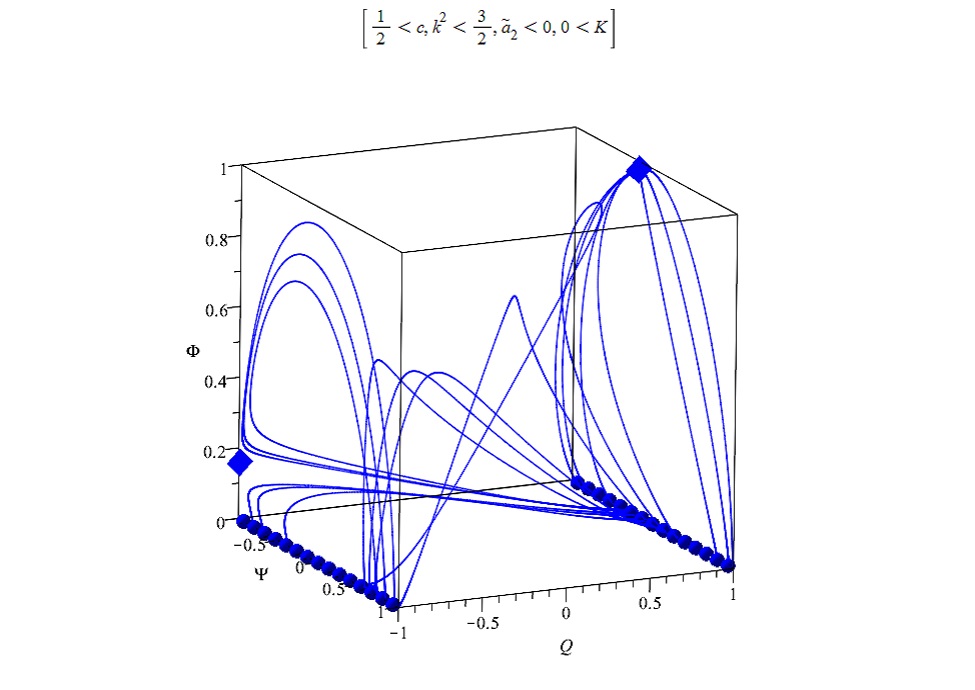}$$
$$\includegraphics[width=0.9\textwidth]{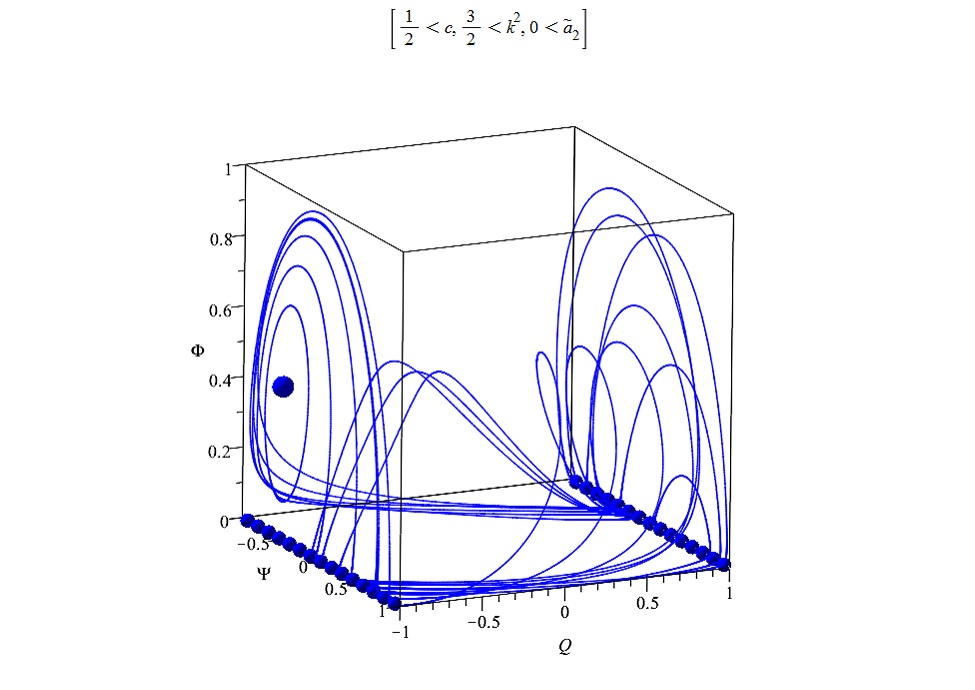}$$
\caption{A selection of phase portraits of the Kantowski-Sachs models in the full phase space.  The small circles represent the line of equilibrium points $C^+$ and $C^{-}$, while the large sphere represents the points $BI_\delta^+$ and $BI_\delta^-$ and diamonds represent the points $FR_\delta^+$ and $FR_\delta^-$.  Note the absence of the two $KS_\delta$ points. Compare with Figure \ref{Phase_Portraits3D1}.} \label{Phase_Portraits3D3}
\end{figure}

\begin{figure}[p]
$$\includegraphics[width=0.9\textwidth]{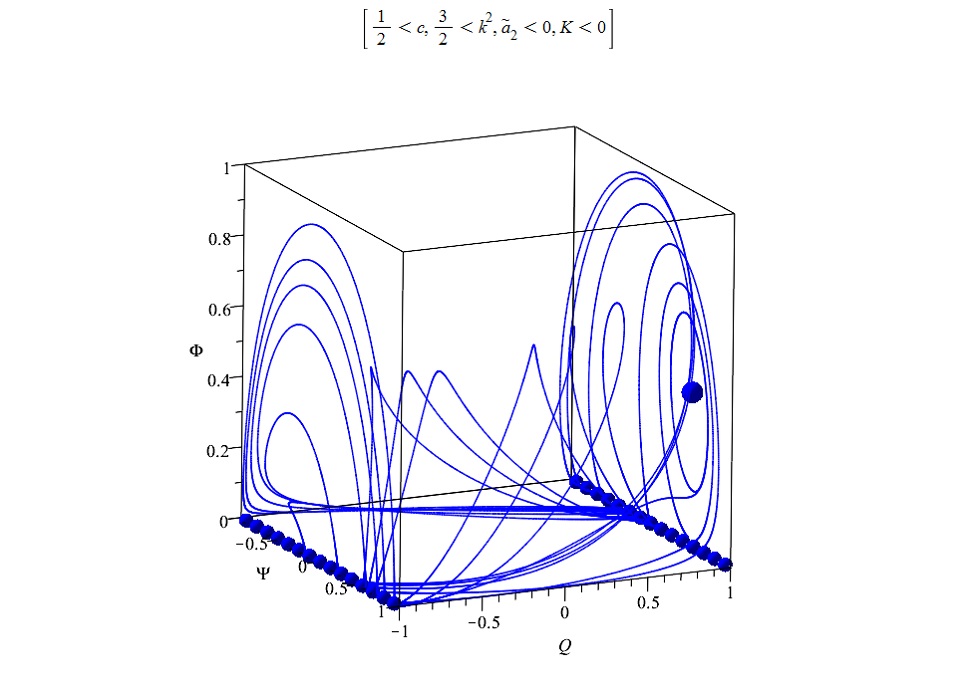}$$
$$\includegraphics[width=0.9\textwidth]{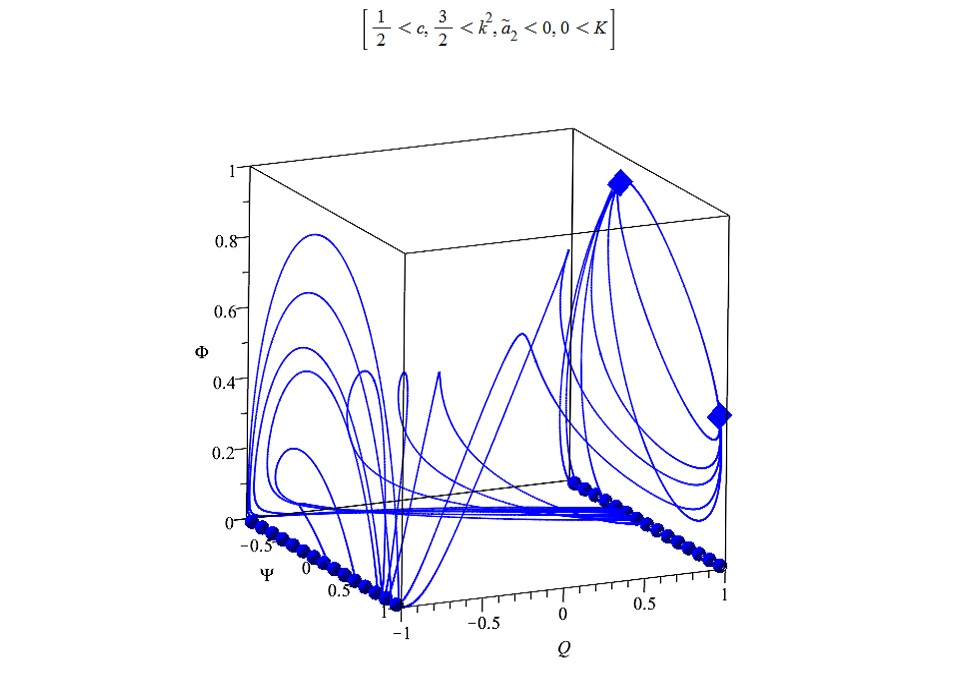}$$
\caption{A selection of phase portraits of the Kantowski-Sachs models in the full phase space.  The small circles represent the line of equilibrium points $C^+$ and $C^{-}$, while the large sphere represents the points $BI_\delta^+$ and $BI_\delta^-$ and diamonds represent the points $FR_\delta^+$ and $FR_\delta^-$.  Note the absence of the two $KS_\delta$ points. Compare with Figure \ref{Phase_Portraits3D2}.} \label{Phase_Portraits3D4}
\end{figure}

There are nine different regions in the $k^2,{\tilde a}_2$ parameter space which represent slightly different dynamical behaviour for the Kantowski-Sachs models.  In addition, we know that $c=\frac{1}{2}$ is also a bifurcation value in the Kantowski-Sachs models, and therefore we need to consider whether $c$ is larger or smaller than $\frac{1}{2}$. The asymptotically stable orbits of the full dimensional dynamical system are summarized in Table \ref{attractor1}.  We observe that the contracting Kasner-like massless scalar field solutions ($C^-$) are stable for all parameter values.  In addition, if $c<\frac{1}{2}$, then the contracting vaccuum Kantowski-Sachs solution, ($KS_-$), is also a stable attractor for all values of $k^2$ and ${\tilde a}_2$.  While for a large range of values of the parameters $k^2$ and ${\tilde a}_2$ we find that there are isotropic, expanding and power-law inflationary solutions acting as attractors, ($FR^+_+$), there are also non-inflationary isotropic attractors, ($FR^-_-$), and even asymptotically stable periodic solutions for a range of parameter values. A selection of phase portraits for the Kantowski-Sachs models are illustrated in Figures \ref{Phase_Portraits3D1}-\ref{Phase_Portraits3D4}.

Similarly, the asymptotically stable orbits in the $Q=+1$ invariant set are also summarised in Table \ref{attractor2}, we note that the dynamics are independent of $c$ in this case.  The asymptotic attractor is either an isotropic, expanding and power-law inflationary solution or the expanding Kasner-like massless scalar field solutions $C^+$. There exists an infinite number of periodic solutions for a range of parameter values.

\begin{figure}[p]
$$\includegraphics[width=0.6\textwidth]{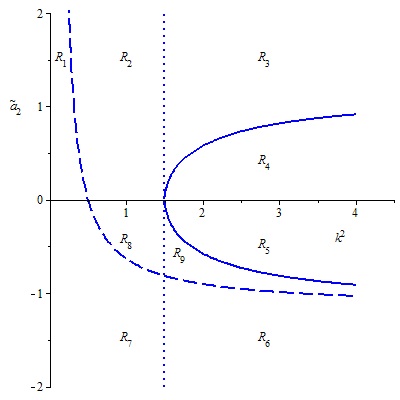} $$
\caption{Parameter space indicating the nine different future asymptotic behaviours possible for the Kantowski-Sachs and Bianchi I models with a coupling between the aether and scalar fields. The solid line is the curve $K=0$, the dotted line is the curve $k^2=\frac{3}{2}$ and the dashed line is the boundary determined by equation \eqref{inflation1_condition}.} \label{Parameter_Space}
\end{figure}

\begin{table}[p]
\renewcommand{\arraystretch}{1.4}
$$\begin{tabular}{|c|l|l|}
\hline
 & \multicolumn{2}{|l|} {Kantowski-Sachs}\\
 \cline{2-3}
Region & $c>\frac{1}{2}$ & $c<\frac{1}{2}$\\
  \hline
  $R_1$ & $C^-$, $FR^+_+$  & $C^-$, $KS^-$, $FR^+_+$ \\
  $R_2$ & $C^-$            & $C^-$, $KS^-$  \\
  $R_3$ & $C^-$, $FR^-_-$  & $C^-$, $KS^-$, $FR^-_-$  \\
  $R_4$ & $C^-$, $BI^-_+$  & $C^-$, $KS^-$, $\left( BI^-_+ \mbox{\ if\ } 4c^2>1-\frac{3}{2k^2}-\frac{3{\tilde a}_2^2}{4}\right)$ \\
  $R_5$ & $C^-$            & $C^-$, $KS^-$ $BI^+_-$ \\
  $R_6$ & $C^-$            & $C^-$, $KS^-$ \\
  $R_7$ & $C^-$, $FR^+_+$  & $C^-$, $KS^-$, $FR^+_+$ \\
  $R_8$ & $C^-$, $FR^+_+$  & $C^-$, $KS^-$, $FR^+_+$ \\
  $R_9$ & $C^-$            & $C^-$, $KS^-$ \\
  \hline
\end{tabular}$$
\caption{List of local attractors for the Kantowski-Sachs models for parameter values in each region $R_i$ given in Figure \ref{Parameter_Space}.} \label{attractor1}
\end{table}

\begin{table}[p]
\renewcommand{\arraystretch}{1.4}
$$\begin{tabular}{|c|l|}
\hline
Region & Expanding Bianchi I\\
  \hline
  $R_1$ & $FR^+_+$           \\
  $R_2$ & $FR^+_+$           \\
  $R_3$ & $C^+$,             \\
  $R_4$ & $C^+$,             \\
  $R_5$ & $C^+$ (Periodic orbits exist) \\
  $R_6$ & $C^+$, $FR^+_+$    \\
  $R_7$ & $C^+$, $FR^+_+$    \\
  $R_8$ & $FR^+_+$           \\
  $R_9$ & $FR^+_+$           \\
  \hline
\end{tabular}$$
\caption{List of local attractors for the expanding Bianchi I models for parameter values in each region $R_i$ given in Figure \ref{Parameter_Space}.}\label{attractor2}
\end{table}

\subsection{Concluding Remarks}

In general relativity with a scalar field having an exponential potential, in order to have an asymptotically stable late-time isotropic and inflationary solution the parameter $k$ is restricted to $k^2<\frac{1}{2}$.  It is interesting to note that the existence of the coupling between the aether and scalar field relaxes this condition and allows for arbitrarily large values of the scalar field potential parameter $k$.  In the models proposed here, the parameter $k$ can be made arbitrarily large as long as one chooses a sufficiently large negative value for the coupling parameter ${\tilde a}_2$.  Indeed, if ${\tilde a}_2<-\frac{2}{\sqrt{3}}$, then any value of $k$ can be chosen.  In this way, the aether field coupling aids in inflation and is reminiscent of assisted inflation using multiple scalar fields \cite{Liddle:1998jc,Malik:1998gy,Copeland:1999cs,Coley:1999mj}.

While the primary purpose of the analysis was to study closed Kantowski-Sachs models, the interesting subcase of expanding Bianchi I models was also analyzed.  The asymptotically stable orbits in this lower dimensional dynamical system are summarized in Table \ref{attractor2}. There we can clearly see how the aether/scalar field coupling changes the dynamics.  If $k^2<\frac{3}{2}$, then all models isotropize to a zero curvature FRW model represented by $FR^+_+$, but do not necessarily inflate unless equation \eqref{inflation1_condition} is satisfied.  The past behaviour is typically a Kasner-like massless scalar field model represented by $C^+$.  If $k^2>\frac{3}{2}$, then a variety of behaviours can exist.  If $k^2>\frac{3}{2}$ and $K>0$, then the isotropic, zero curvature FRW model represented by $FR^+_+$ is the late time attractor (again not necessarily inflationary unless equation \eqref{inflation1_condition} is satisfied) while a massless scalar field model $C_+$ and the isotropic $FR^+_-$ are the sources.  If $k^2>\frac{3}{2}$ and $K<0$,  then the late time attractor are massless scalar field models represented by some portion of $C_+$ while the other portion of $C_+$ is the source, but interestingly enough, there also exists periodic orbits in this case.

We note that for a particular range of parameter values, there exists a family of asymptotically stable period orbits.  These orbits can exist for both expanding and contracting models. These families of periodic orbits exist in the Bianchi I invariant sets and therefore can be stable late-time attractors classes of anisotropic but spatially homogeneous models with a scalar field with an exponential potential and a aether field/scalar field coupling.


\acknowledgments RvdH thanks the Department of Mathematics and Statistics at Dalhousie University for their kind hospitality.  RvdH is supported by the St. Francis Xavier University Council on Research.  AAC is supported by the Natural Sciences and Engineering Research Council of Canada. BA would also like to thank the Government of Saudi Arabia for financial support.

\providecommand{\href}[2]{#2}\begingroup\raggedright\endgroup

\end{document}